\documentclass[10pt, onecolumn]{IEEEtran}
\usepackage{array}
\usepackage{subfigure}
\usepackage{xcolor, colortbl}
\usepackage{hyperref}
\usepackage{url}
\usepackage{booktabs, tabularx, colortbl,threeparttable}
\usepackage{multirow,amsmath,mathrsfs,verbatim, bbm}
\usepackage{graphicx}
\usepackage{authblk}
\definecolor{mygray}{gray}{.7}
\usepackage{listings}
\usepackage{xspace}
\usepackage{caption2}

\newcommand{\nickName}{AppStarscream\xspace}

\begin{document}
\bibliographystyle{abbrv}

\title{Flexible Installability of Android Apps with App-level Virtualization based Decomposition}

\author{Yi Liu}
\author{Yun Ma}
\author{Xuanzhe Liu}
\affil{}
\renewcommand\Authfont{\fontsize{10}{11.4}\selectfont}
\renewcommand\Affilfont{\fontsize{9}{10.8}\itshape}

\maketitle

\begin{abstract}
With the popularity of smartphones, users are heavily dependent on mobile applications for daily work and entertainments. However, mobile apps are becoming more and more complicated with more features and increasing size, part of which may be redundant to users. Due to the limitation of current installation mechanism, users have to download full-size applications instead of enjoy only the wanted features. Such full-size apps may consume more resources, including CPU, memory, and energy, which may hurt users' enthusiasm for further installation. We first conduct an empirical study to characterize used features when users interact with mobile applications, and find that users only consume a small set features of target apps. To address this problem, we present \nickName{}, which offers to decompose and run Android apps with app-level virtualization. With \nickName{}, developers can decompose an existing app into multiple bundles, including a base bundle with frequently used pages and feature bundles with inactive pages. Then, end users can just download base bundle for daily use, and visit uninstalled pages on demand. We have implemented a prototype system and evaluated it with real-world apps showing that \nickName{} is efficient and practical.
\end{abstract}

%
% The code below should be generated by the tool at
% http://dl.acm.org/ccs.cfm
% Please copy and paste the code instead of the example below.
%
%\begin{CCSXML}
%<ccs2012>
% <concept>
%  <concept_id>10010520.10010553.10010562</concept_id>
%  <concept_desc>Computer systems organization~Embedded systems</concept_desc>
%  <concept_significance>500</concept_significance>
% </concept>
% <concept>
%  <concept_id>10010520.10010575.10010755</concept_id>
%  <concept_desc>Computer systems organization~Redundancy</concept_desc>
%  <concept_significance>300</concept_significance>
% </concept>
% <concept>
%  <concept_id>10010520.10010553.10010554</concept_id>
%  <concept_desc>Computer systems organization~Robotics</concept_desc>
%  <concept_significance>100</concept_significance>
% </concept>
% <concept>
%  <concept_id>10003033.10003083.10003095</concept_id>
%  <concept_desc>Networks~Network reliability</concept_desc>
%  <concept_significance>100</concept_significance>
% </concept>
%</ccs2012>
%\end{CCSXML}
%
%\ccsdesc[500]{Computer systems organization~Embedded systems}
%\ccsdesc[300]{Computer systems organization~Redundancy}
%\ccsdesc{Computer systems organization~Robotics}
%\ccsdesc[100]{Networks~Network reliability}

\begin{IEEEkeywords}
	App decomposition, App-level virtualization, On-demand installation
\end{IEEEkeywords}

\IEEEpeerreviewmaketitle

\section{Introduction}
\label{sec:introduction}

In the past decade, we have witnessed the burst of mobile applications (a.k.a., apps). Apps have played an indispensable role in our daily life and work, and even changed the way that we interact with the external world. People become to download the \textit{installation packages} from app stores such as Apple AppStore and Google Play, and perform various tasks such as web browsing, social networking, watching online videos, playing games, and so on.

Millions of software developers have gained great success on the appstore-centric ecosystem, e.g., rewarded by thousands of downloads, high ranking, and even revenues. However, it is also true that a large number of developers are not lucky enough and their apps never win the expected success. One recent study over millions of Android users indicates that the popularity of apps typically complies with the power law, i.e., only a small portion of apps account for substantial downloads and user-interaction time~\cite{Liu:TSE17}\cite{WWW15Li}. It is also reported that more than 90\% of apps are launched for only once after they are downloaded~\cite{WWW15Li}. Although there could be various reasons why users tend to dislike and uninstall some apps, it should be arguable whether \textit{these ``abandoned'' apps are ``completely'' useless to users}. It is possible that some of the features of these apps are still useful to some ``\textit{opportunistic and situational}'' requirement. For example, a tourist may need to search for local news on ``\texttt{Lyon Daily}'', and compare the expense of renting the same car model via various e-car-rental platforms. Apparently, she needs only some relevant features of these apps but the current app-distribution model requires her to download and install apps. Therefore, she is likely to give up downloading the apps and the developers' opportunity of winning this user is compromised.

By contrast, it is also true that we may not access every single feature of even those frequently used apps. One fact is that current apps become very complex and account for large-volume installation packages. For example, the most popular app in China, \texttt{WeChat}, now contains more than 600 ``activities''\footnote{The basic program unit on  Android OS. Suppose that an app can be analogous to a website, then an activity refers to a webpage of this site. We exchangeably use ``page'' and ``activity'' in the rest of this paper.} along with thousands of features, and stands for more than 40 MB package size. In fact, such a problem is known as the ``\textit{software bloat problem}''~\cite{software_bloat} of Android, and many users complain that the increasing package size not only requires local storage, but also leads to the sluggish experience due to more local computing-resource consumption and background collusion~\cite{WWW17Xu}.

An ideal way to mitigate the software bloat shall be that every single page of an app can be installed in a ``service-oriented'' fashion at the page level rather than downloading the whole package as Web apps~\cite{ICWS15Liu}. Such a desire is reasonable due to many successful Software-as-a-Service applications such as Salesforce and Google Docs. In practice, we notice that some solutions are stepping toward this way. For example, the recent Google Instant apps~\cite{instantapp} provide an ``installation-free'' style to Android users by clicking through a hyperlink to access a desired page of an app, rather than downloading this app. However, Google Instant apps require the system-level support: they can be deployed on only versions after Android 5.0 and require the developers' manual efforts to grant the API level higher than level-21~\cite{instantapp}. Tencent WeChat announces its Mini Program platform~\cite{miniprogram}, which works in the similar way. These Mini Programs are essentially customized Web applications implementing some features that are thus accessed via hyperlinks from the app. Unfortunately, Mini Programs suffer from poor user experience compared to native apps.

However, given that current apps are usually developed in an iterative style and incrementally updated over the code base from their previous versions, it is challenging  to re-design and re-develop millions of existing on-the-shelf apps to be service-oriented. Therefore, we need to rethink how to make apps better deployed and delivered.  We broadly require that apps should be refactored to equip the following abilities: (1) \textit{\textbf{decomposable}}, i.e., the ability that an app can be decoupled into relatively independent pages; (2) \textit{\textbf{deployable}}, i.e., the ability that end-users can easily and regularly deploy the apps on devices without additional requirements compared to the current app-installation mechanism; (3) \textit{\textbf{user-friendly}}, i.e., the ability that the apps should operate as expected and smoothly without compromising the user experience; (4) \textit{\textbf{developer-friendly}}, i.e., the ability that developers can transform apps to be service-oriented without re-development.

In this paper, we propose a novel approach, called \nickName{}, to help developers decompose and re-deploy existing Android apps for on-demand and flexible installability. 
\nickName{}  aims to meet all the four aforementioned requirements. 
\emph{Decomposable}: \nickName \emph{decomposes} an Android app into a core ``\emph{base bundle}'' and the remaining pages as installable features that can be loaded in an on-demand way. A base bundle contains frequently used ``pages'' and the dependencies promising the reachability of every single page and the correct behavior of the whole app, including code and resources. In this way, developers can then choose to release the app with only the base bundle and a list of pages that can be selected by users according to their personal preferences. 
\emph{Deployable}: \nickName{} provides an app-layer virtualized controller that resides between the app and the underlying OS, which is a normal Android app that users can download and install. 
\emph{User-friendly}: When users access pages that have not been installed, \nickName{}'s runtime execution environment dynamically requests these pages from a remote cloud where a full version is deployed, and guarantees original functionalities and user interface.
\emph{Developer-friendly}: \nickName provides a fully automated approach that leverages static analysis to identify what pages and resources form the base bundle, and includes an iterative and back-complementary recovery mechanism with recording and replay to supplement the code and resources missed by the static analysis. 
As such, apps can be automatically transformed without involving re-development from developers.

This paper makes the following main contributions:

\begin{itemize}
	\item We motivate our research based on an empirical study and reveal that the page-level app installation and access are non-trivially  required.
	\item We design a virtualized app-layer controller to properly decompose an Android app for more flexibly personalized and customized, while  preserving the correct functionality and smooth user experience.
	\item We demonstrate that our approach is practical by decomposing 1,000 real-world Android apps and evaluate the performance after decomposition along with on-demand installation. Our approach can save \textbf{44.17\%} initial download size, and reduce \textbf{10.9\%} and \textbf{20\%} time on launching base bundles and feature bundles in median case, respectively. Meanwhile, our approach can reduce \textbf{6.5\%} and \textbf{10.7\%} memory usage when launching base bundles and feature bundles in the median case, respectively.
\end{itemize}

\section{A Motivating Empirical Study}
\label{sec:motivation}

First, we aim to know how and how much an app is used at a fine granularity inside an app. To this end, we conduct an empirical study to understand the evolving app complexity and app usage patterns in Android apps by answering three research questions:

\begin{itemize}
	\item {\textbf{RQ 1:} \textit{How does the functionality complexity of apps evolve as the apps upgrade?} As apps keep upgrading, they tend to become more and more complex with more functionalities to attract users. By answering this question, we can understand the software bloat problem in the Android ecosystem.}
	\item {\textbf{RQ 2:} \textit{How many features in an app are used by different users?} Although mobile apps tend to provide more features to retain their users, users usually use only certain features of an app rather than all of them, just as the Pareto principle~\cite{Pareto_principle} points out. By answering this question, we can understand whether all provided features of an app are really necessary for users.}
	\item {\textbf{RQ 3:} \textit{Do different users focus on different features in an app?} By answering this question, we can understand the diversity of app usage patterns.}
\end{itemize}

\begin{figure}
   \centering
  \subfigure[The APK volume of the first version and the latest version]{
    \label{fig:topapp:a}
    \includegraphics[width=0.4\textwidth]{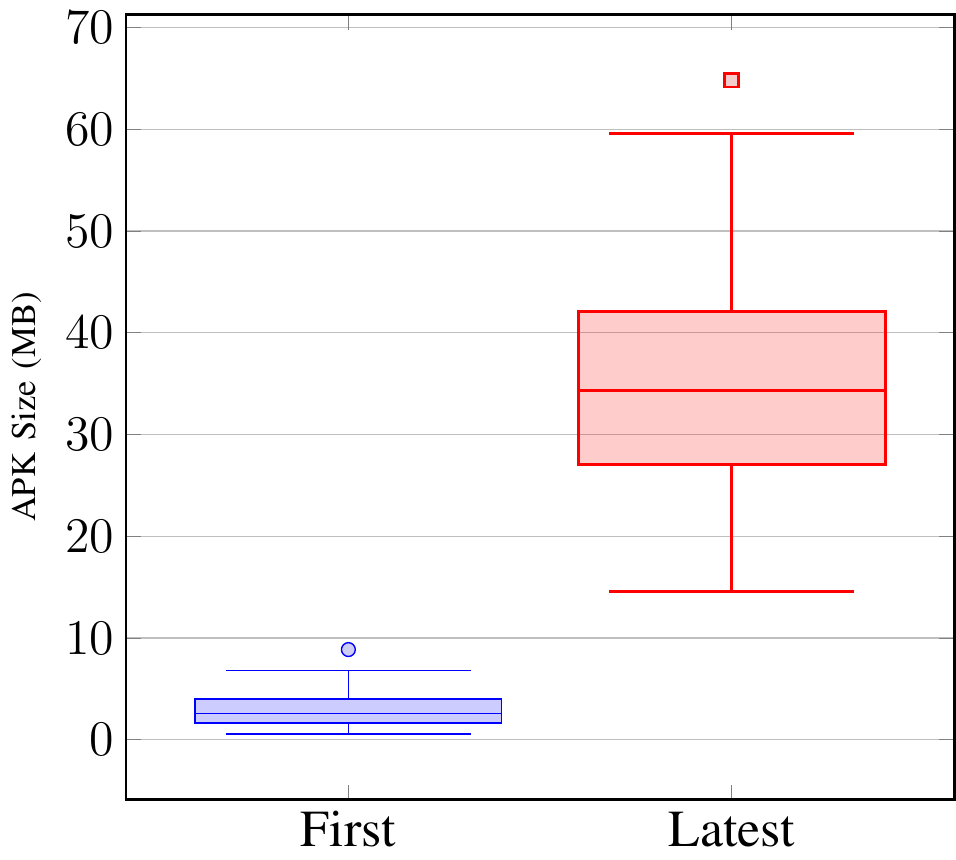}}
  \hspace{0in}
  \subfigure[The activity number of the first version and latest version]{
    \label{fig:topapp:b}
    \includegraphics[width=0.44\textwidth]{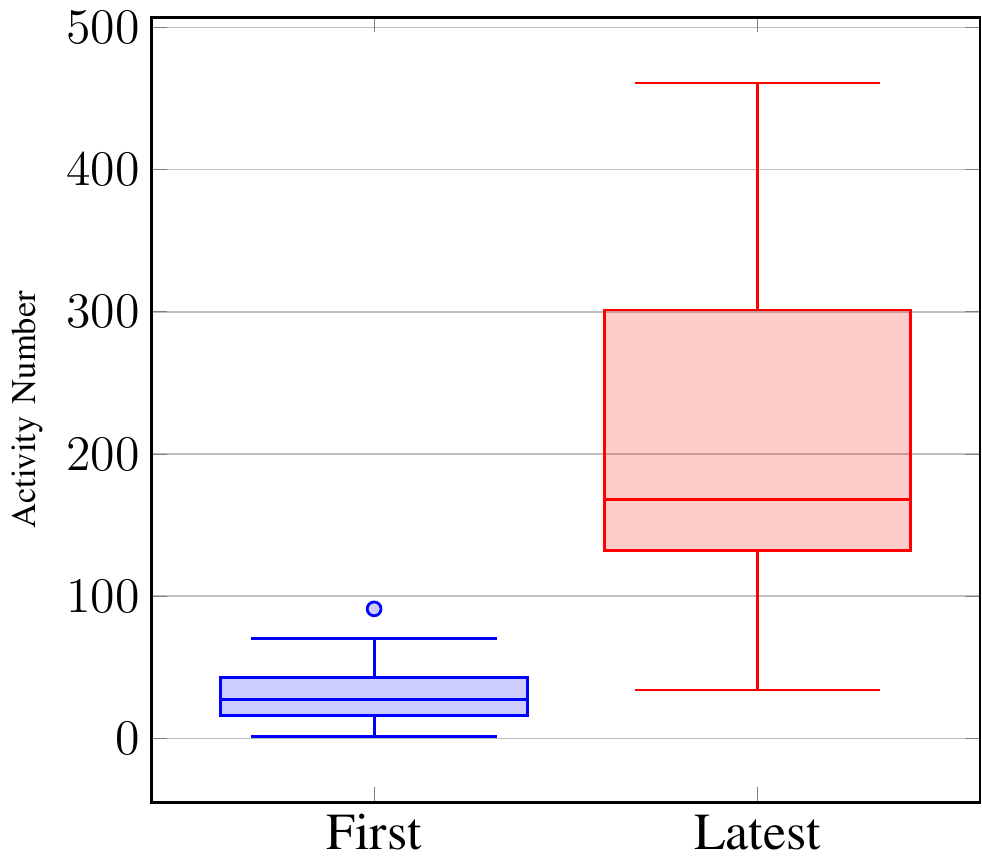}}
  \caption{The change of APK size and activity number of top 50 apps in Wandoujia market}
  \label{fig:topapp}
\end{figure}

\subsection{Ever-Increasing App Complexity}
To study RQ 1, i.e., \emph{how does the functionality complexity of apps evolve as the apps upgrade}, we choose the top-50 most downloaded apps in a leading Android AppStore in China, namely  Wandoujia\footnote{Till 2017, Wandoujia has more than 4 millions of users and 1.8 millions of apps~\cite{wandoujia}}.
For simplicity, we download each app's APK file (which is the installation package of Android apps)  for the first version that can be publicly found and the latest version in October, 2017.
We use the size of APK files and the number of activities as two metrics to measure an app's functionality complexity because (1) larger installation packages usually imply more features, and (2)  activities, also referred to pages, are the basic components in Android apps to provide a unique interactive functionality for users.

Figure~\ref{fig:topapp} shows the distribution of APK size and the number of activities of the two versions from the 50 chosen apps.
We find that both the APK size and the number of activities have increased dramatically, by comparing the first version and the latest version.
With the version evolving, apps have increased by 3.16$\sim$60.76 times in size and 1.53$\sim$143 times in the number of activities.
The median values for the size increase and the number of activities are 13.85 times and 8.44, respectively.
Such results indicate that Android apps are becoming more and more complex with the apps evolve.

Take WeChat\footnote{WeChat is a popular IM app with more than 900 millions of users.} as an example. When WeChat was first released in 2011, it just provided the basic feature of instant messaging, and the size of installation package was just 4.74MB. However, after 6 years, the latest version of WeChat provides a large number of features, such as online shopping, online payment, and news reading. Meanwhile, the size of WeChat's installation package has increased to 40.42MB, about 8.5 times larger compared to the first version.

\subsection{Sparse App Usage}

To study RQ 2, i.e., \emph{how many features in an app are used by different users}, we conducted a field study to understand how users interact with mobile apps. We developed an Android client, which periodically records the activity name and the package name of the foreground app. Then, we hired 38 in-college student volunteers and deployed the client on their Android smartphones. The data collection lasted for three months (Feb to April, 2016), and we finally collected 894,542 records, containing 3,527 activities from 389 apps\footnote{The data collection and analysis process was conducted with strict IRB approval from the Research Ethic Committee of the authors' institute that is anonymous for review.}. We filter out the records related to system apps and self-developed apps that cannot be downloaded in Android markets, and finally acquire a dataset consisting of 240 apps.

\begin{figure}
   \centering
  \subfigure[Users only visit part of features in mobile application]{
    \label{fig:used_activity_rate}
    \includegraphics[width=0.4\textwidth]{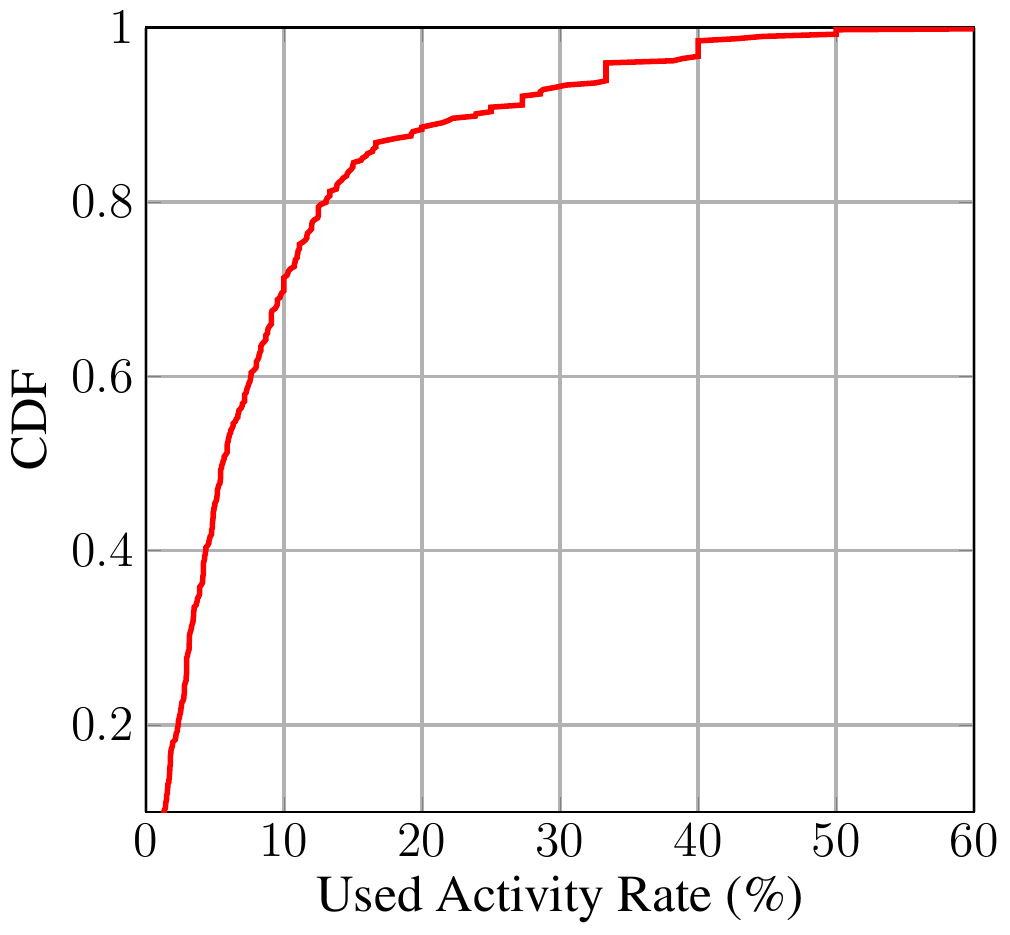}}
  \hspace{0in}
  \subfigure[The entropy of visited pages for each application]{
    \label{fig:visited_pages_entropy}
    \includegraphics[width=0.4\textwidth]{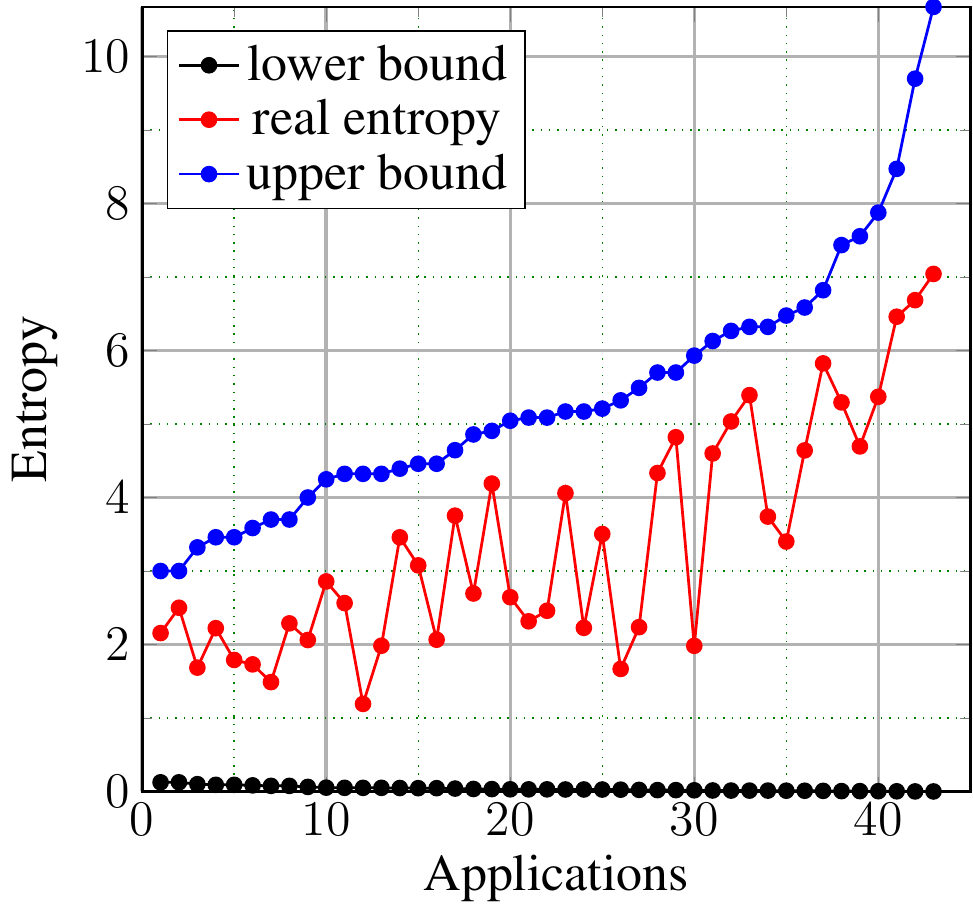}}
  \caption{An empirical study of how end users interact with mobile applications}
  \label{fig:empirical_study}
\end{figure}

For each app, we count the number of all the unique activities that are visited by the app's users, and divide the number by the total number of activities, to compute the feature-usage ratio of the app. Figure~\ref{fig:used_activity_rate} depicts the distribution of the feature-usage ratio among all the 240 apps. We can see that the users visit only less than 20\% of the activities in most of the apps (about 87\%). Such a result indicates that the users need only a small set of features in mobile apps but they have to download the full-size installation package. Meanwhile, Li~\cite{WWW2016Li} finds that about 60\% of abandoned apps can survive for only less than 1.5 days, and about 80\% of abandoned apps can survive for less than a week. From the users' perspective, it is annoying to make unnecessary features in their devices, resulting in more occupation of local storage, memory, and other resources. As such, an infrequently used app of huge size could be removed by the user in a short time.

\subsection{Diverse Usage among Users}
To study RQ 3, i.e., \emph{do different users focus on different features in an app}, we still use the dataset collected for RQ 2, choose the apps that have more than 5 users, and obtain 43 apps in total. We simply use the entropy~\cite{entropy} to measure the diversity of activity usage among users. We count the number of users for each visited activity of an app, and each visited activity has a probability of being visited by a user. For every single app, we can calculate the entropy as shown in Equation~\ref{entropy},  where $n$ is the number of activities that have been visited by users, and $p_{i}$ is the probability of being visited for an activity.
\setlength{\belowdisplayskip}{0pt} \setlength{\belowdisplayshortskip}{0pt}
\setlength{\abovedisplayskip}{0pt} \setlength{\abovedisplayshortskip}{0pt}
\begin{equation}
	\label{entropy}
	E=-\sum _{i=1}^{n}p_{i}\ln p_{i}
\end{equation}

If all activities are visited equally, all $p_{i}$ values equal $1 / n$, and the entropy hence takes the value $\ln(n)$ (upper bound). The more unequal probabilities of being visited, the larger the weighted geometric mean of the $p_{i}$ values, and the smaller the corresponding entropy. If users' visits focus on one activity, and other activities are rarely visited (even if there are many of them), entropy approaches zero (lower bound). Figure~\ref{fig:visited_pages_entropy} shows the distribution of the entropy among the 43 apps. We can find that each application has an entropy between its upper bound and lower bound, indicating that its activities have quite diverse probability of being visited.

\subsection{Findings and Implications}

Our preceding empirical study shows that Android apps become more and more complex with increasing size and activities. However, only a limited set of activities are frequently visited, and some activities may never be visited. This finding motivates us to \emph{remove those infrequently visited activities} to mitigate the software bloat problem in Android ecosystem.

However, we also observe that different users focus on different features in an app.  This observation implies that these users may need to use an infrequently visited activity occasionally. Therefore, we cannot just simply remove those infrequently visited activities, which would cause crashes when the users visit the removed activities.
We need to devise a mechanism that allows users to not only keep \emph{the basic and frequently used features} when the apps are downloaded,
but also can visit \emph{the infrequently used features whenever necessary} (i.e., on demand).

\section{Requirements and Key Challenges}
Motivated by the empirical study, we propose an approach that aims to alleviate the software-bloat problem in Android apps.
Android app development advocates the component-based development paradigm, and thus app developers tend to design their apps to be modularized.
Such modularity of Android apps present opportunities for \emph{decomposing an app into separated bundles}, each of which contains parts of the app's code and resources.
For each app, users just need to download and install a base bundle that can correctly launch the app and include the frequently used features.
In this way, the users keep only those code and resources related to their desired features on the devices so that the device's storage could be saved and the app's performance such as memory usage could be improved.
When users need to use the features not in the base bundle (i.e., those infrequently used features), we could allow the related bundles to be dynamically downloaded and installed at runtime,
achieving \emph{on-demand installation of bundles}.
Considering the current Android ecosystem, the proposed approach should satisfy the following requirements:

\noindent\textbf{Decomposable}. The approach should decompose an app into several bundles. Each bundle should be correctly launched and used just as the original app does.

\noindent\textbf{Deployable}. Users can easily and regularly deploy the decomposed apps on their devices without additional requirements compared to the current app-delivery mechanism. If any modifications of the Andoird system were requested, it is impractical to deploy and distribute our approach.

\noindent\textbf{User-friendly}. The approach should not compromise the user experiences or change the way of user interactions. The decomposed app should behave as expected as if the original app is being used.

\noindent\textbf{Developer-friendly}. The approach should work for legacy and on-the-shelf apps so that developers do not have to manually re-develop their apps to adopt the approach.

However, there are two key challenges for Android app decomposition into bundles and on-demand installation of the bundles.
First, it is difficult to obtain the complete dependencies among code and resources via only static analysis. For example, developers may use Java reflection to invoke methods or access resources indirectly. If we cannot decompose an app correctly, it would suffer crashes due to missing code or resources.
Second, it is difficult to enable the on-demand installation without modifying the Android system. We need to dynamically load the newly-downloaded code and resources without changing Android's runtime. Meanwhile, we need to identify the exact entry point to launch the feature activity and dynamically load feature bundles beforehand. However, in some cases, the target activity can be decided only by the OS at runtime if developers use an implicit intent to launch the target activity.
For example, an app may use string concatenation based on variables to construct activity names. Without correctly resolving the activity name in runtime, it is impossible to dynamically load feature bundles beforehand.

\section{AppStarscream}
\label{sec:approach}

This section describes how \nickName{} works and the detailed design of each component.
\subsection{Approach Overview}

\begin{figure}[t]
	\centering
    \includegraphics[width=0.6\textwidth]{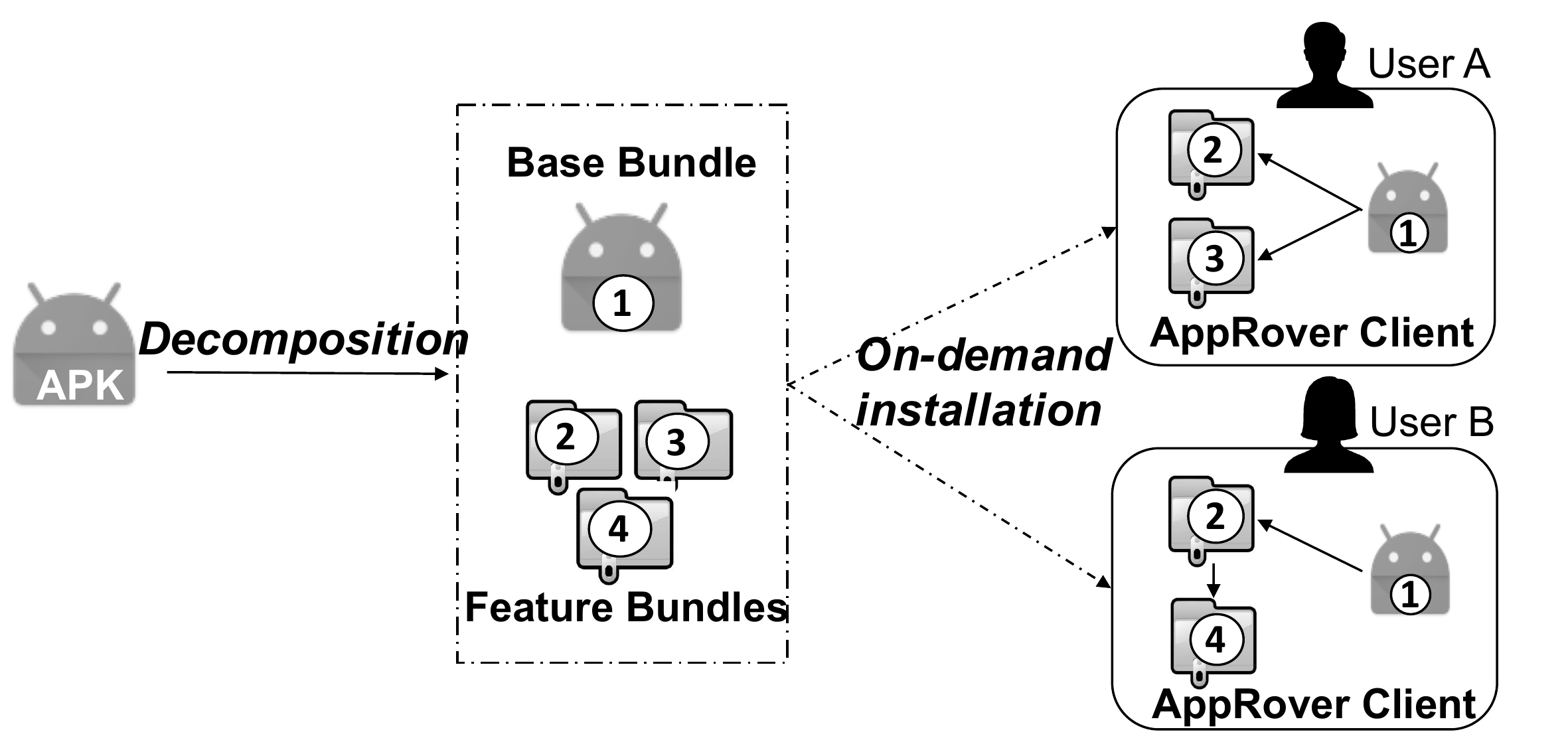}
    \caption{Approach Overview} 
    \label{fig:approach_overview}
\end{figure}

Figure~\ref{fig:approach_overview} shows the overview of \nickName{}. Given an Android app, \nickName{} automatically decomposes the app into decoupled bundles, and enables users to install these bundles on demand. The decomposition of \nickName{} performs at the activity level, i.e., each activity is regarded as an atomic feature that may be desired by users~\cite{Ubicomp17Lu}. We choose this decomposition level because Android advocates component-based development and developers tend to design different features in different activities, which are Android's basic components that provide user interfaces for users to interact with. After decomposition, \nickName{} generates two types of bundles:

\begin{itemize}
	\item {A \textbf{base bundle}, which contains the code and resources related to app launching and the frequently used
activities. In general, launching an app is to load its main activity. But before loading the main activity, there may be some other \emph{``welcome''} activities, such as splash activities to show advertisements. Therefore, \nickName{} always keeps the code and resources related to the main activity as well as those \emph{welcome} activities into the base bundle. In addition, base bundle also includes code and resources that cannot be decomposed and should be mandatorily included in the installation package such as Services, BroadcastReceivers, and assets, which are important to ensure the correctness of the app behaviors.}
    \item {A set of \textbf{feature bundles}, each of which contains the code and resources related to a specific activity (called feature activity) that
        is not in the base bundle. A feature bundle is requested when users navigate to a new activity that is not included in the base bundle and has not been visited yet. By default, all feature bundles are not installed on the device. When users' actions trigger the navigation to an unvisited activity, AppRover downloads the code and resources from the corresponding feature bundle, merges resources, and loads the code dynamically.}
\end{itemize}

In order to support on-demand installation of bundles, \nickName{} has an app-level virtualization space to hook system services and take over the whole life-cycle management of an app, including installation, running, and uninstallation\footnote{In theory, it can be integrated into the Android system as system services to achieve better performance, but it would require modifications of the Android system.}. The app-level virtualization space is hosted as an Android app called \emph{\nickName{} client}. Users just need to deploy the \nickName{} client on their devices, and choose to install and launch the decomposed apps (actually are the corresponding base bundles). To some extent, we can regard the \nickName{} client as a Web browser: the Web browser downloads Web-page related resources and renders Web pages on demand; the \nickName{} client downloads and installs feature bundles on demand. The difference is that users should manually download the base bundle of an app to launch it in the \nickName{} client, unlike Web browser where all the Web pages are visited in the same way.
In the remaining part of this section, we describe the details of app decomposition, app virutalization, and on-demand installation, respectively.

\subsection{App Decomposition}

\begin{figure}[!tbp]
    \includegraphics[width=0.95\textwidth]{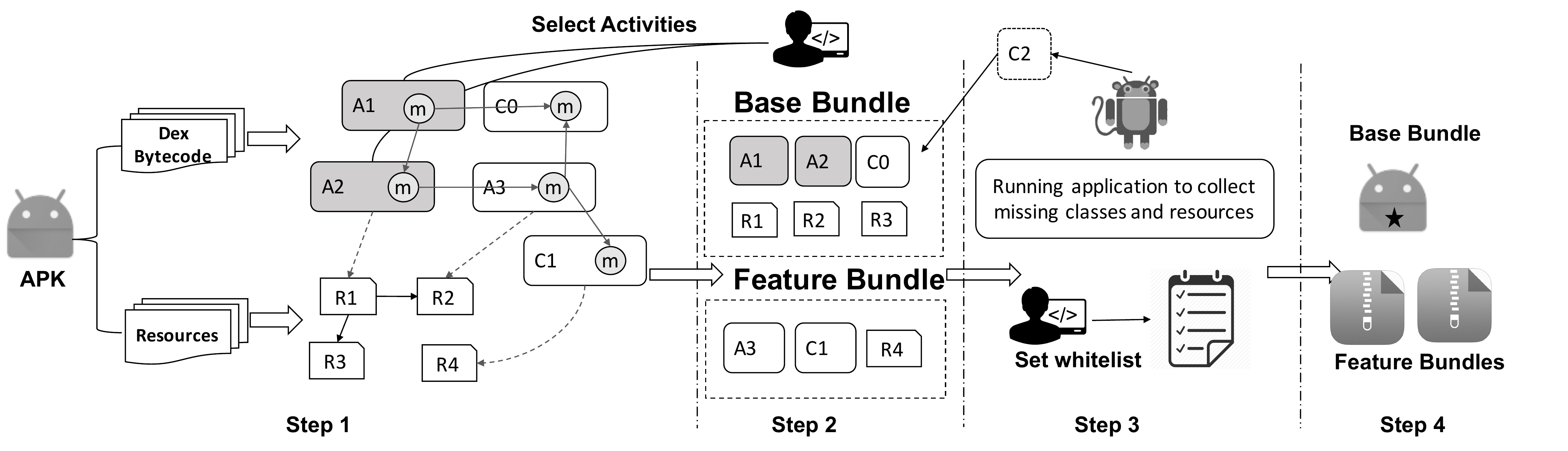}
    \caption {App Decomposition Workflow}
    \label{fig:workflow}
\end{figure}

We formally define the terms that are used to describe our approach for app decomposition.

\noindent\textbf{Basics of Android App}.
We first present the definition of an Android app, which is the input of \nickName.
We define $APP=<CLS, RES, OTH>$ to represent an app, where $APP$ is a triple including $CLS$ (i.e., the compiled code), $RES$ (i.e., resources), and $OTH$ (i.e., other resources). 
We define $CLS= ACT\cup OLS \cup POJO$, where $CLS$  is the set of classes in bytecode, including $ACT$ (activities), $OLS$ (other Android components), and $POJO$ (other classes). 
We define $RES=IDR\cup ASSET$, where $RES$ represents the resources of an app, including $IDR$ (resources that reside in \textit{res} folder, and can be referenced with identifiers such as \textit{R.type.resName} in code and \textit{@type/resName} in resource files) and $ASSET$ (resources that reside in \textit{assets} folder, and can be referenced with \textit{AssetManager} object and \textit{loadUrl} method). The former reference can be detected explicitly, while the latter reference can not be detected explicitly\footnote{For example, developers may load a local Web page in \textit{assets} folder with \textit{loadUrl} method in a WebView component. However, the loading of other resources such as CSS files, JS files, and images in \textit{assets} folder is not explicitly specified.}. Therefore, we focus on decomposition of resources in the \textit{res} folder. 
$OLS$ represents other Android components that include services, broadcast receivers, and content providers.
The $OTH$ represents other resources in an APK file that we will pack into the base bundle directly, such as native libraries, \textit{AndroidManifest} file, and \textit{resources.arsc} file.

We define the relationships between code and resources as $Refer\subset CLS\times IDR$ , where $<c,r> \in Refer$ if a class $c\in CLS$ references a resource $r\in IDR$. We build a dependency graph $RG=<IDR, F>$ among resources from $IDR$, in which each node $r_i\in IDR$ denotes a resource and an edge $e(r_i,r_j)\in F$ denotes that $r_i$ references $r_j$. Similarly, $path_{r_i,r_j} \in Path_{RG}$ denotes t!hat there exists a reference path $P=r_ir_1r_2\dots r_j$ such that $<r_i,r_1>\in F$ and for $t=1,2,\dots,j, 1\le j, <r_{t-1}, r_t>\in F$. 

For each class $c_k \in CLS$, its related resources can be defined as $CRRES_{c_k}=\{r\in IDR|<c_k, r>\in Refer \lor \exists r_t\in IDR, <c_k, r_t>\in Refer \land <r_t, r>\in Path_{RG}\}$. 
Each $r\in CRRES_{c_k}$ is either a resource that is directly related to $c_k$ (i.e., $<c_k,r>\in Refer$), or a resource that is referenced by another resource $r_t$ related to $c_k$ (i.e., $<c_k, r_t>\in Refer$ and $<r_t, r>\in Path_{RG}$).

\noindent\textbf{Base Bundle and Feature Bundle}.
Based on the definition of an Android app, we can give the formal definitions of the base bundle and the feature bundle as shown in Definition~\ref{base_bundle} and Definition~\ref{feature_bundle}. 
A base bundle consists of code ($CLS_{base}$), referenced resources ($RES_{base}$), and others ($OTH$). 
Developers can choose a set of activities defined as $SEL=\{a_1, a_2, \dots, a_n\}$ to pack into the base bundle
Based on the chosen activities, we compute the set of \emph{activity-related classes} $ARCLS_{a_i}$.
We consider that a class $c$ such that $c\in POJO$ is required by an activity $ACT$ if there exists a calling relationship between the methods in the $ACT$ and the methods in $c$.
We define that a class consists of a set of methods as $CLS=\{m_i|m_i\in METHOD\}$. 
We define the call graph as $CG=<METHOD, E>$, in which each node $m_i\in METHOD$ denotes a method and an edge $e(m_i,m_j)\in E$ denotes a calling relationship from $m_j$ to $m_j$. $path_{m_i,m_j} \in Path_{CG}$ denotes that there exists a call path $P=m_im_1m_2\dots m_j$ such that $<m_i,m_1>\in E$ and for $t=1,2,\dots,j, 1\le j, <m_{t-1}, m_t>\in E$. For each activity ${a_k}$, we define $ARCLS_{a_k}=a_k\cup\{c\in POJO|\exists m_i\in a_k, m_j\in c, <m_i, m_j>\in Path_{CG}\}$, where $ARCLS_{a_k}$ is the activity-related classes of  ${a_k}$, including the ${a_k}$ and class $c$ such that $c\in POJO$ and $m_i\in a_k, m_j\in c, <m_i, m_j>\in Path_{CG}$. 

Based on activity-related classes, $CLS_{base}$ is defined as the union of $ARCLS_{a_i}$ for $a_i \in SEL$ and the other Android components $OLS$. 
We preserve all the components in $OLS$ since they do not have related resources and usually occupy a small amount of storage.
$RES_{base}$ is defined as the union of class related resources $CRRES_{c_i}$ for each $c_i$ in $CLS_{base}$ and the assets $ASSET$. For each activity $a_k$ thatis  not included in base bundle, we pack its classes $CLS_{feature_{a_k}}$ and resources $RES_{feature_{a_k}}$ as a feature bundle $FEATURE\_BUNDLE_{a_k}$. $CLS_{feature_{a_k}}$ denotes the set of related classes in $ARCLS_{a_k}$ except those classes in $CLS_{base}$. $RES_{feature_{a_k}}$ denotes the set of referenced resources that are in the union of $CRRES_{c_i}$ for each class $c_i$ in $CLS_{feature_{a_k}}$ but not in the base bundle's referenced resources $RES_{base}$.
\setlength{\belowdisplayskip}{0pt} \setlength{\belowdisplayshortskip}{0pt}
\setlength{\abovedisplayskip}{0pt} \setlength{\abovedisplayshortskip}{0pt}
\begin{subequations}
\small
\label{base_bundle}
\begin{align}
	&BASE\_BUNDLE = <CLS_{base}, RES_{base}, OTH> \\
	&CLS_{base} = \{\bigcup_{i=1}^{n} ARCLS_{a_i}\}\cup OLS \\
	&RES_{base} = \{\bigcup_{c_i\in CLS_{base}} CRRES_{c_i}\}\cup ASSET
\end{align}
\end{subequations}
\begin{subequations}
\small
\label{feature_bundle}
\begin{align}
	&FEATURE\_BUNDLE_{a_k} = <CLS_{feature_{a_k}}, RES_{feature_{a_k}}> \\
	&CLS_{feature_{a_k}}  = ARCLS_{a_k}-CLS_{base} \\
	&RES_{feature_{a_k}} = \{\bigcup_{c_i\in CLS_{feature_{a_k}}} CRRES_{c_i}\}-RES_{base}
\end{align}
\end{subequations}

\noindent\textbf{Overall Workflow and A Working Example}. 
Figure~\ref{fig:workflow} illustrates the workflow of app decomposition. The input is an APK file $APP$.
Our approach first identifies the resources $RES$ and builds the call graph $CG$ of the code $CLS$ by static analysis. 
For each class $c_i$ in $CLS$, our approach computes the relationships among code and resources based on $CG$ and obtains the referenced resources $CRRES_{c_i}$.
Second, based on the set of activities specified by the developers, our approach decomposes the app into a base bundle and multiple feature bundles as we defined before. 
Figure~\ref{fig:workflow} shows an example of how our approach decomposes an app. Assume that the developers choose the activities $A1$ and $A2$ to pack into the base bundle. 
Then, the activity-related classes ($ARCLS_{A1}$ and $ARCLS_{A2}$) and their related resources ($CRRES_{A1}$ and $CRRES_{A2}$) are computed based on $CG$.
Based on these results, the POJO class $C0$ and the resources $R1$-$R3$ are found to be required by the activities $A1$ and $A2$ and will be packed into the base bundle. 
Since $C0$ and $R2$ have been packed into the base bundle, we will not pack them into the feature bundle that contains the activity $A3$.
Third, since the static analysis based $CG$ may not find all the code and resources required by the chosen activities, our approach further applies an iterative and back-complementary recovery mechanism with recording and replay to supplement missing code and resources for the base bundle and feature bundles. 
During this step, our approach detects that the running base bundle throws out an exception, which indicates the class $C2$ is missing. As such, our approach adds $C2$ back to the base bundle, and repeats this process iteratively until no exceptions are detected. We also allow developers to manually add some classes and resources to a white list.
At last, we repack these selected activities into a base bundle with their dependent code and resources. Similarly, other activities will be packed as feature bundles, respectively.

\noindent\textbf{Iterative and Back-Complementary Recovery Mechanism}. Our approach requires a precise and complete call graph, which is very challenging due to the event-driven nature and virtual calls in Android apps~\cite{Wang16PLDI}\cite{Li17IST}\cite{Li16ISTTA}. 
To alleviate such issue, we devise an iterative and back-complementary recovery mechanism with recording and replay to supplement missing code and resources for the base bundle and feature bundles.

When running a base bundle without certain required classes and resources, a system exception indicates that the required classes or resources cannot be found will be thrown.
Our approach analyzes such exception information to infer missing classes and resources. 
Our approach provides a record-replay tool and a back-complementary recovery tool to automatically complement missing code and resources, repack bundles, launch target activities, and back and forth. 
The back-complementary recovery tool iteratively installs and launches the base bundle, and adds the detected missing code and resources back to base bundle until successfully launching the home page. 
For each feature activity, the developers need to record actions based on the record-replay tool to reach the target activity, such as launching the application, clicking a specific button. 
Then, the back-complementary recovery tool iteratively replays the recorded actions to open the target activity, and collects the logs about missing code and resources. 
Additionally, developers can also manually specify code and resources for the base bundle and feature bundles in a white list if they want. 
This can help reduce the iterations if developers are familiar with the app.

\subsection{App-level Virtualization}

\begin{figure}[!t]
	\centering
    \includegraphics[width=0.7\textwidth]{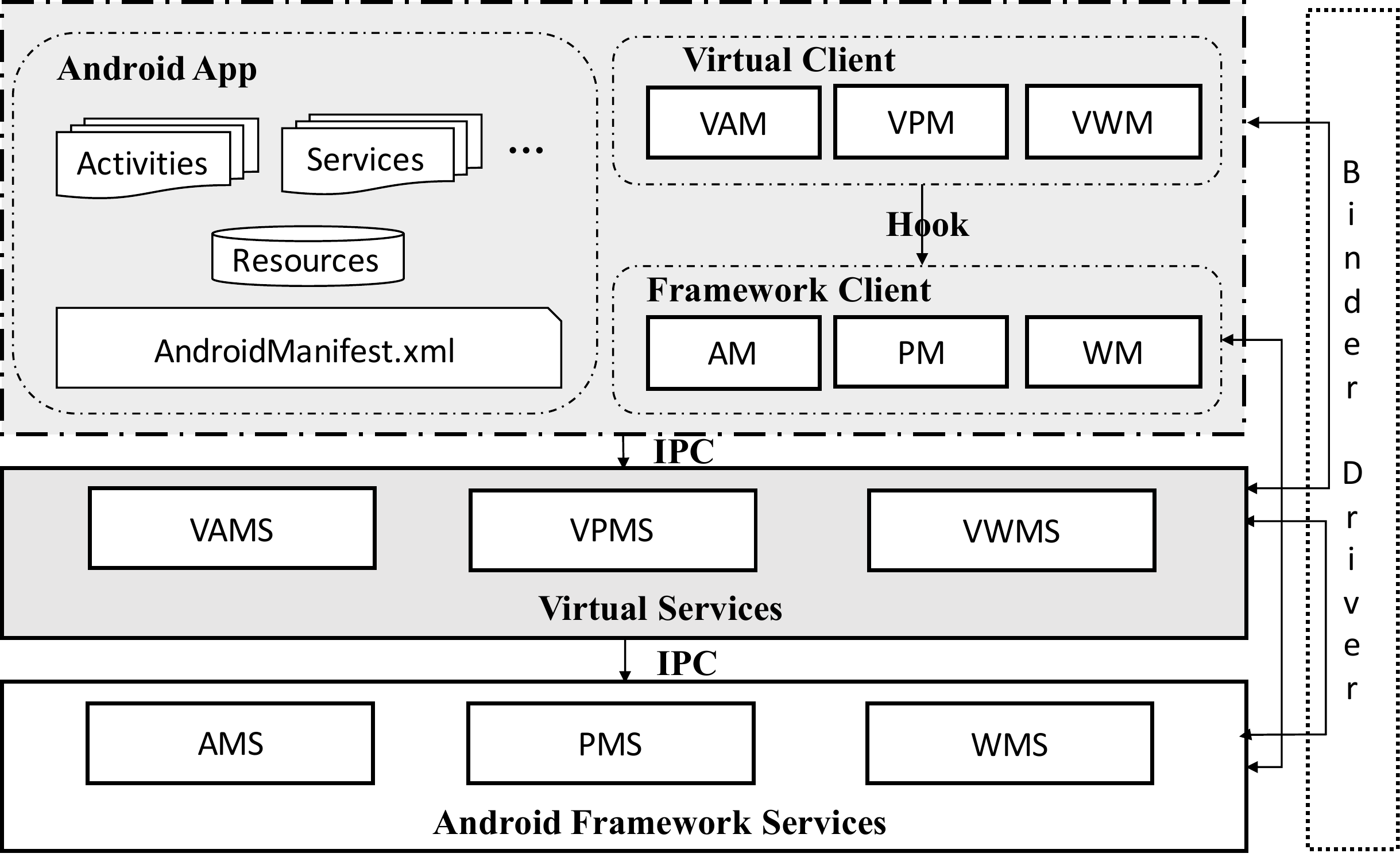}
    \caption{Architecture of App-level virtualization} 
    \label{fig:app_level_virtualization}
\end{figure}
\nickName{} provides an app-level virtualization space to take over the management of the whole lifecycle of a decomposed app, including installation and runtime support. The key challenge is to hook the related system services, including \textit{ActivityManagerServiche} (AMS, responsible for managing lifecycles of activities), \textit{PackageManagerService} (PMS, responsible for installation and package info management), and \textit{WindowManagerService} (WMS, responsible for window management, events management and distribution) to deal the whole lifecycle of an Android app at runtime. Although we can use many kinds of existing technologies, such as Xposed~\cite{xposed}, AspectJ~\cite{MP2009Laddad}, CydiaSubstrate~\cite{CydiaSubstrate}, these technologies require either low-level accesses or modifications to the underlying Android OS, resulting in limited deployability and robustness for the decomposed apps.

Android system manages and runs apps with the aforementioned system services via the Binder~\cite{binder}, which is an Android-specific inter-process communication mechanism and remote method invocation system. 
It adopts a client-server model, where each client would have a proxy object to communicate with the server.  
We find that app processes and system services act as both clients and servers to deal with the lifecycle management by cooperation. 
For an Android app, its process holds the clients of system services (e.g., the activity manger, short as AM, is the client of AMS). We can hook these client proxies to deceive system services.

Figure~\ref{fig:app_level_virtualization} shows the architecture of our app-level virtualization framework, \emph{\nickName{} Client}. The \nickName{} client provides a server process that maintains a set of fake services, including VAMS, VPMS, and VWMS corresponding to AMS, PMS, and WMS in the Android OS, respectively. After downloading the base bundle of an app in \nickName{} client, the VPMS will extract and parse the \textit{AndroidManifest.xml} as the PMS does when installing a regular application directly. 
Additionally, we need to address the issue that the Android OS disallow to run an activity that is not registered in the \textit{AndroidManifest.xml} file. 
Hence, \nickName{} client registers many empty stub activities, and invokes the exact lifecycle callbacks of the target activity when running the decomposed apps. 

When end users launch the base bundle, the \nickName{} client forks a new process to load the code of base bundle and inquires the exact launching activity through VPMS. Decomposed apps running in the \nickName{} client are not aware of the existence of the \nickName{} client as if it is running directly in the Android OS. In fact, they interact with our virtualized services, including VAMS, VPMS, and VWMS through the respective hooked virtual clients, including VAM, VPM, and VWM. Then, the \nickName{} client intercepts the message that tells the VAMS to start an activity and pass the info of a stub activity previously registered to AMS. When AMS tries to invoke the callbacks of the stub activity, the \nickName client replaces the stub activity with the exact launching activity of the base bundle to deceive AMS into opening the launching activity.

\subsection{On-Demand Installation}
\nickName allows end users to download only the base bundle of a decomposed app when they install a new application. 
The \nickName{} client downloads the base bundle, and parses it to obtain the app information as normal installation in the Android OS.
For those explicit intents that will open uninstalled activities, we inject a piece of code by bytecode rewriting~\cite{OOPSLA2012Zhang} before the statements. For those implicit intent, the VPMS will resolve the intent object to find the matched activity. When the end users' operations trigger the navigation to a new activity, the \nickName{} client detects if the activity has resided in the local device. If not, the \nickName{} client initiates a request to download the feature bundle that contains the target activity. After the download, the \nickName{} client extracts all the resources and merges them into the base bundle. Then, the \nickName{} client extracts the code and moves it to the private source code folder of the app so that the application can dynamically load the code with dynamic code loading (DCL) based on \textit{DexClassLoader}.  With both the code and resources dynamically loaded, the application can now navigate to the target activity. Every requested feature bundle is downloaded once, and subsequent visits on the activities in these bundles will load the activities from the local device directly, improving delays and user experiences.

\section{Implementation}

According to the design in Section~\ref{sec:approach}, we have implemented a prototype for \nickName{}.
We build a static analysis tool based on Soot~\cite{soot} to generate call graphs and perform bytecode rewriting for Android apps.  
Our static analysis tool also inspects the code to find the identifiers that reference resources (e.g., in the format of \textit{R.type.name}) and computes the relationships between the code and the resources.
To analyze resource files, we build another tool based on Apktool~\cite{apktool} to decode resources from APK files and analyze the dependencies among resources. 
 
Our iterative and back-complementary recovery tool is implemented upon the \textit{Android Debug Bridge (ADB)} tool and the \textit{MonkeyRunner}~\cite{monkeyrunner}.
The tool leverages the ADB tool to search apps' logs for entries that contain \textit{ClassNotFoundException}, so that \nickName can detect the missing code. 

Our record and replay tool is implemented based on the MonkeyRunner, and developers can record a sequence of actions to launch the target activity. 
We implement the \nickName{} client to enable virtualization space based on VirtualApp~\cite{virtualapp}, which is an open-source Android container.
\section{Evaluation}
\label{sec:evaluation}

To evaluate the effectiveness of \nickName, we have conducted comprehensive evaluations from three aspects. 
First, we apply \nickName on an extensive set of apps to demonstrate its effectiveness in saving the initial download size of an Android app and the download/installation time. 
Second, we evaluate the robustness of the iterative back-complementary recovery tool on a set of open-source apps. 
At last, we evaluate the runtime performance of the decomposed apps in the \nickName{} client.

\subsection{Applicability over Real-World Apps}
In order to evaluate our practicality of real-world apps, we first analyze 1,000 apps downloaded from Google Play to measure the size distribution of base bundles and feature bundles.

\begin{figure}[!t]
   \centering
  \subfigure[Base bundle]{
     \label{fig:size_of_base_bundle}
    \includegraphics[width=0.4\textwidth]{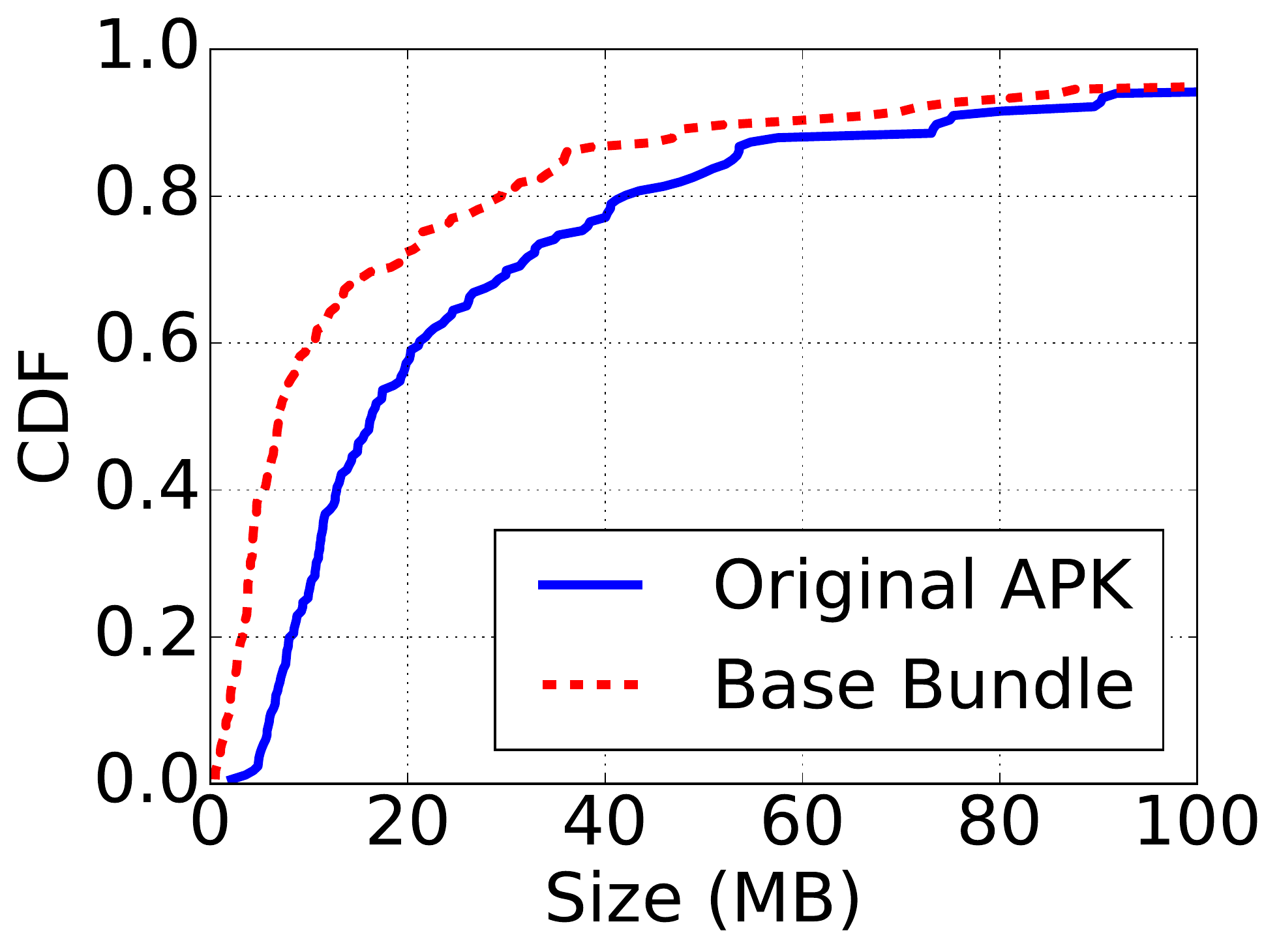}}
      \hspace{0in}
   \subfigure[Feature bundle]{
     \label{fig:size_of_feature_bundle}
    \includegraphics[width=0.4\textwidth]{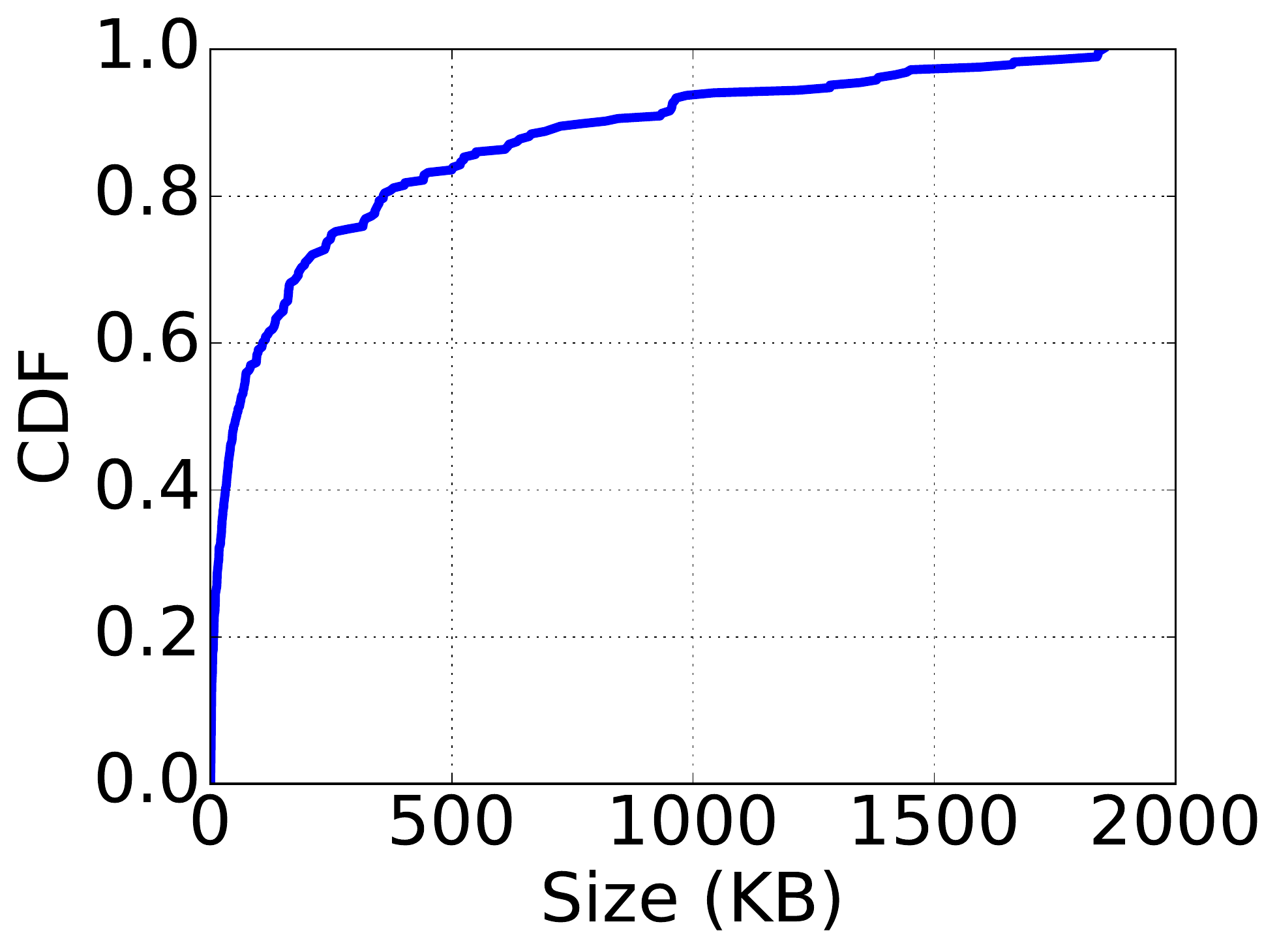}}
      \hspace{0in}
      \label{fig:decomposition}
      \caption{Distribution of the size of decomposed bundles}
\end{figure}

\noindent\textbf{The size of base bundles}. Figure~\ref{fig:size_of_base_bundle} shows the distribution of  size of base bundles and their original apps.
The results show that the median value for the saving of the size of the initial download is \textbf{44.17\%}, resulting in a significantly shorter download time and less local storage.

For some apps, the base bundles do not save much storage compared to original apps. 
The reason is that the developers may place most of the features in the lauching activity of their apps, and most of the other activities' dependent resources and code have been packed into base bundles. Meanwhile, our current implementation just decomposes resources that reside in the \textit{res} folder, and other resources are packed into the base bundle directly. Some apps place almost all of resources in the \textit{assets} folder (e.g., mobile games), so the savings of storage for such apps are much less.

\noindent\textbf{The size of feature bundles}. As shown in Figure~\ref{fig:size_of_feature_bundle}, users need to download only a feature bundle whose size is less than 500 KB for about \textbf{84.23\%} cases. To make a comparison, a recent study of \textit{httparchive.org}~\cite{httparchive} reports that the average web page size in 2016 was 2, 232 KB. In other words, a feature bundle is much smaller than a common Web page in size, and thus end users do not need to wait a long time to download it. The reason is that some feature bundles may just contain a small set of classes, and their dependent resources have been packed into the base bundle. Therefore, their size is relatively small.

\nickName can lower the barrier for accessing a new mobile app because of a smaller-size base bundle, especially for those big-size apps. Meanwhile, users just need to download a small-size feature bundle to visit a new activity that has not been downloaded on their devices.

\subsection{Robustness of Decomposed Apps}
\label{subsec:robustness}

\begin{figure}[!t]
\centering
\begin{minipage}[t]{0.4\linewidth}
	\centering
   	\includegraphics[width=1\textwidth]{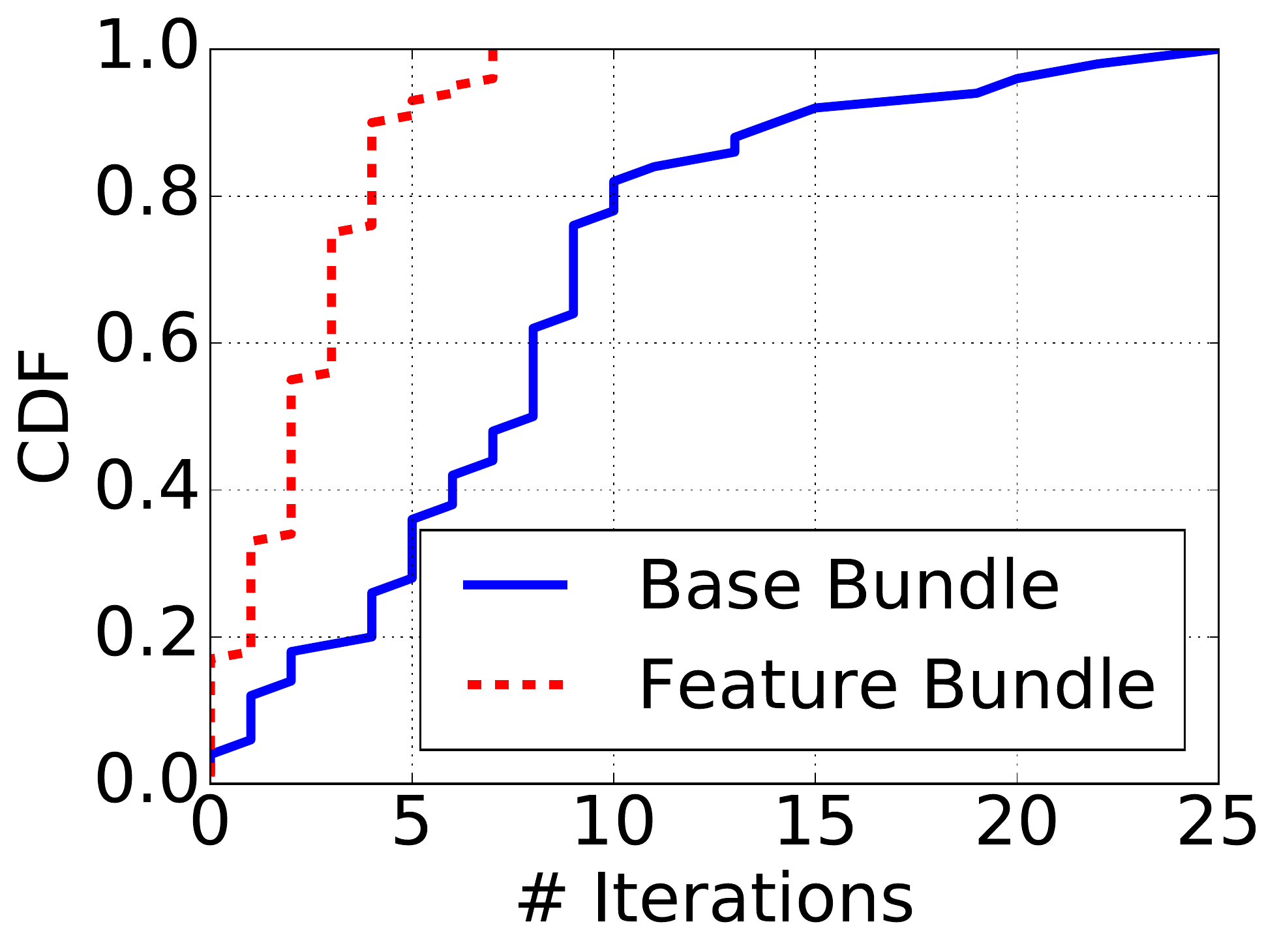}
   	\caption{Distribution of the iteration times for bundles}
   	\label{fig:iterative_count}
\end{minipage}\hspace{.1cm}
\begin{minipage}[t]{0.4\linewidth}
	\centering
   	\includegraphics[width=1\textwidth]{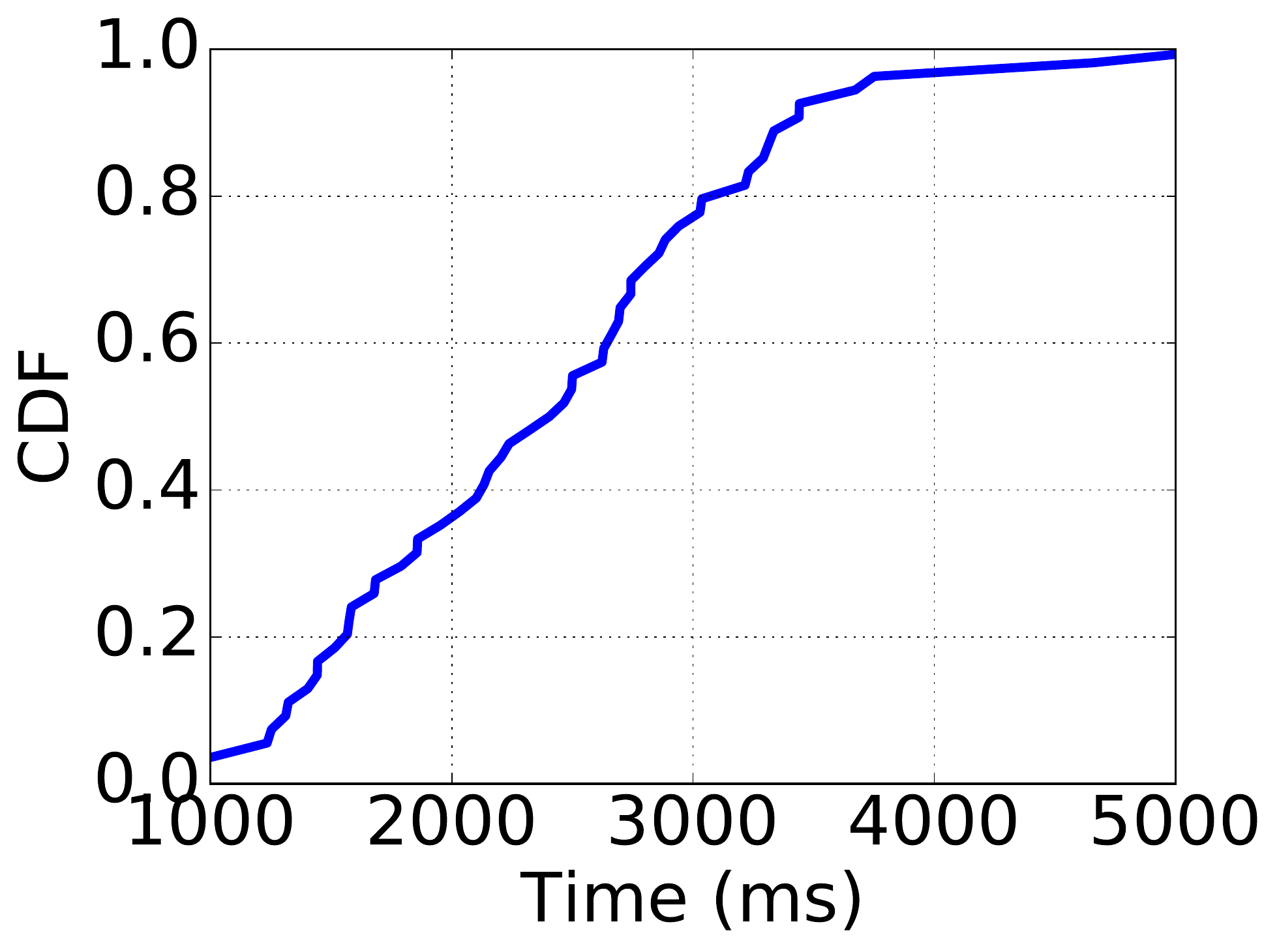}
    \caption{Distribution of the time spent on merging feature bundles}
    \label{fig:time_on_merge_feature_bundle}
\end{minipage}
\end{figure}

To evaluate the robustness of the iterative back-complementary recovery mechanism, we conduct an evaluation on a set of open-source apps.
Although we can conduct evaluations on APK files directly, developers may have used some tools to protect their app from being reversely engineered. 
Therefore, we could fail to run the decomposed apps. 
To avoid such issue, we download 50 open-sourced apps from the F-droid\footnote{F-Droid is an installable catalogue of FOSS (Free and Open Source Software) apps for the Android platform (\url{https://f-droid.org/})} and Github and verify the correctness of our decomposition process.

For each app, we choose the home activity and those \emph{welcome} activities to generate the base bundle, and choose activities that are successor nodes of the home activity on the activity transition graph~\cite{OOPSLA13Azim} to generate feature bundles. We inject a piece of code to directly return before the code that opens an activity not included in either the base bundle or the feature bundles.

\noindent \textbf{Effectiveness and Efficiency of Supplementing Missing Code and Resources}.  We use the iterative back-complementary recovery mechanism to supplement the missing code and resources so that the decomposed apps can run correctly. We count the required iteration times to supplement the base bundle and the feature bundles for the 50 apps. Figure~\ref{fig:iterative_count} shows that we can successfully supplement the missing code and resources after 8 iterations, and no more than 10 iterations for 80\% of the apps. Notice that every iteration needs to add only one class. Compared to the number of classes (ranging from 1160 to 5817, 3347 in the median case) in the base bundle, the additional cost of iterations is quite marginal. Meanwhile, we find that supplementing feature bundles usually needs fewer iterations. This is because most of the missing code and resources in a feature bundle have been already included in the base bundle. We also find that many apps use some popular third-party libraries, and thus they miss same classes. In future work, we plan to use this finding to optimize the efficiency of the mechanism when we detect apps use these libraries.

\noindent \textbf{Runtime Correctness}. We further verify the decomposition correctness of \nickName{} on the 50 open-source apps. 
We first use the Monkey~\cite{monkey} tool to generate random streams of user events, such as clicks, touches, or gestures, as well as a number of system-level events, to run the original application for one minute. 
During the executions, we use the MonkeyRunner~\cite{monkeyrunner} to record these actions.
The recorded actions are then used to replay the decomposed apps running in the \nickName{} client for a fair comparison. 
For each app, we run both the original app and the decomposed app 10 times to collect the logs. 
We successfully run our decomposed apps in the \nickName{} client without crashes. 
According to the collected logs, our decomposed apps do not throw extra exceptions or cause errors during the executions, demonstrating the robustness of our approach.

\subsection{User-Perceived Performance}

\begin{figure*}[!t]
   \centering
  	\subfigure[Launching time of base bundles]{
     \label{fig:time_on_launch}
    \includegraphics[width=0.4\textwidth]{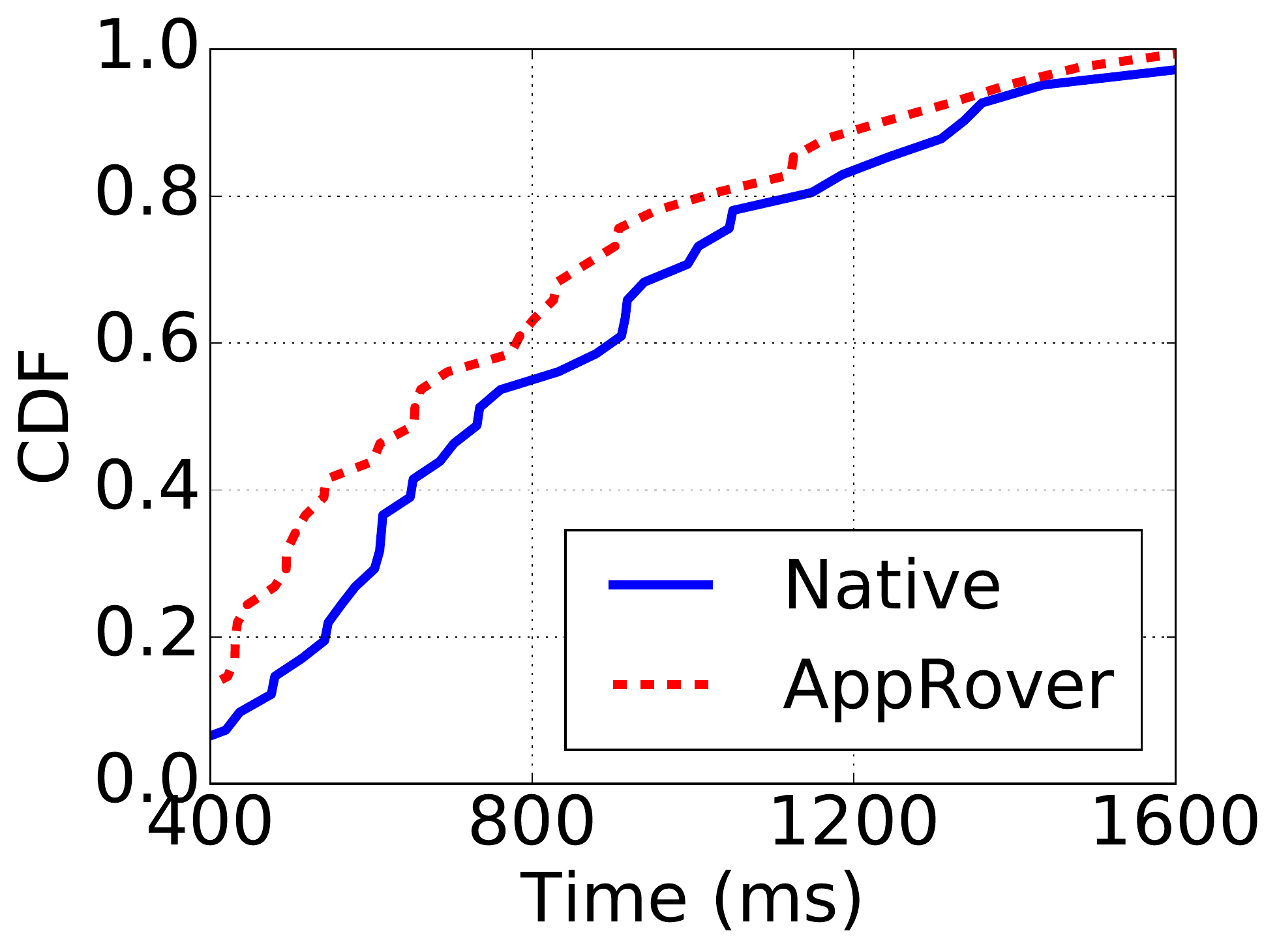}}
      %\hspace{0in}
    \subfigure[Launching time of feature bundles in the warm start]{
    \label{fig:feature_load_time_warm}
    \includegraphics[width=0.4\textwidth]{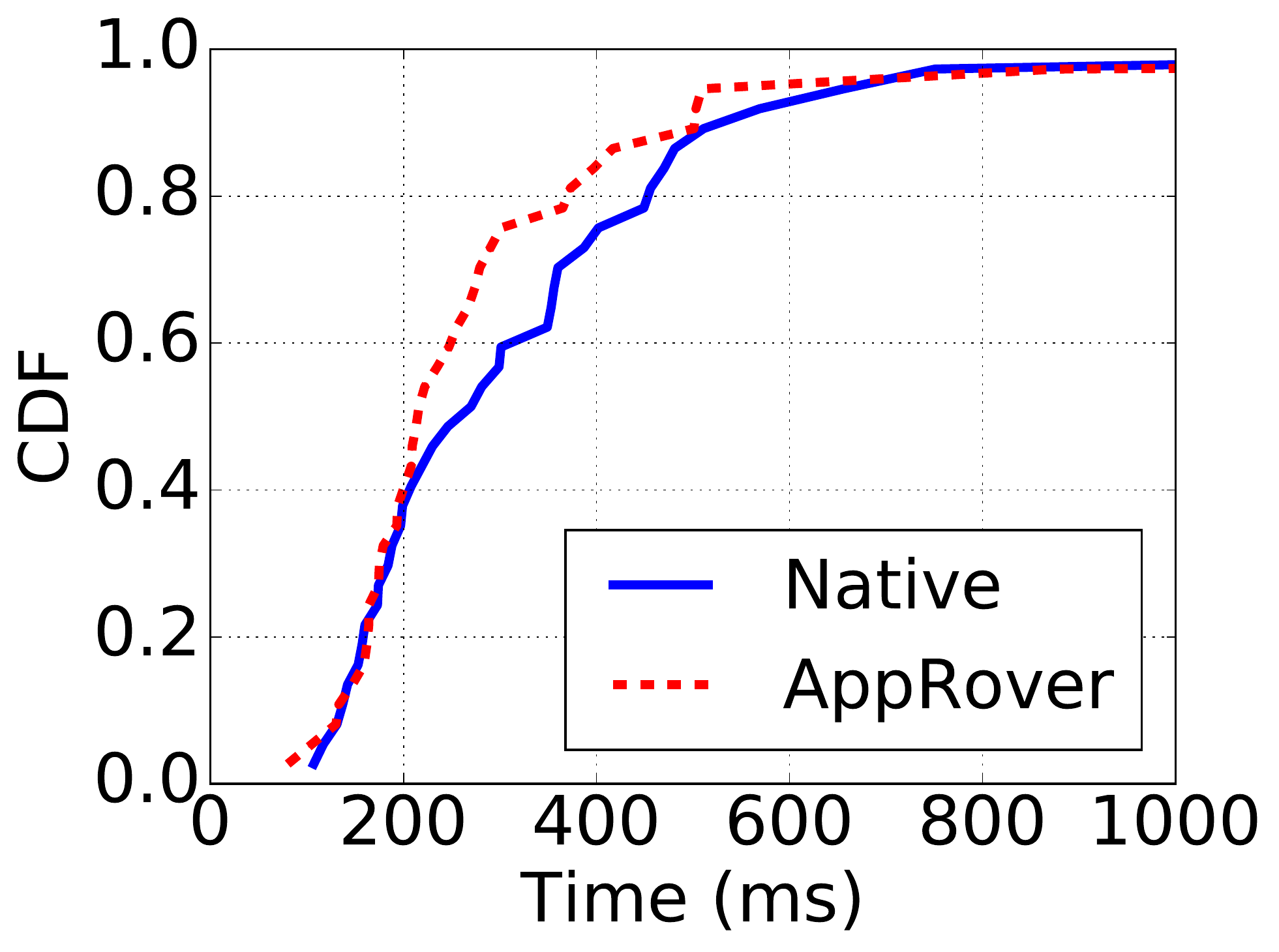}}
  	%\hspace{0in}
  	\subfigure[Memory usage of launching base bundles]{
    \label{fig:launch_memory}
    \includegraphics[width=0.4\textwidth]{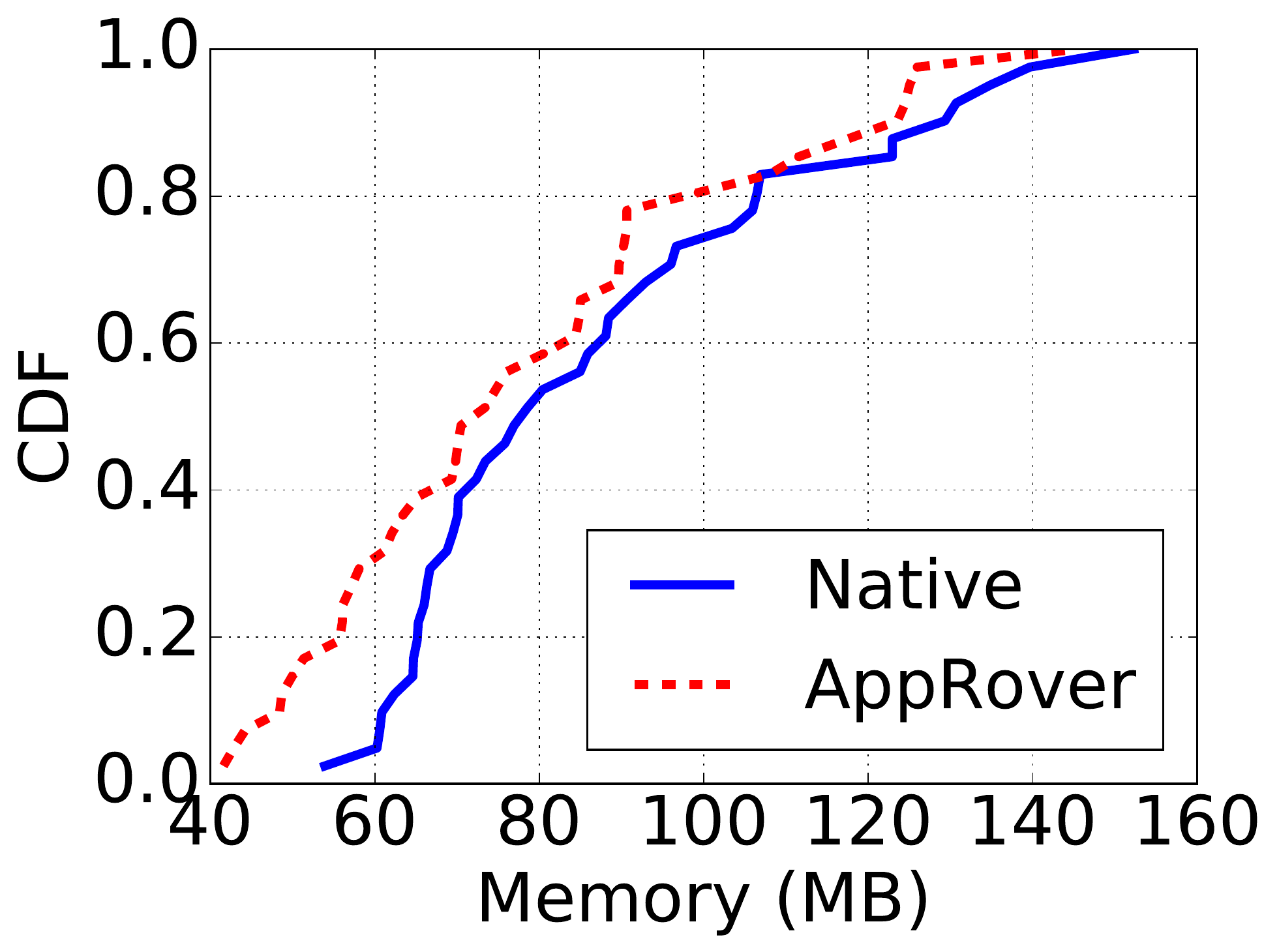}}
     % \hspace{0in}
    \subfigure[Memory usage of launching feature bundles]{
    \label{fig:load_memory}
    \includegraphics[width=0.4\textwidth]{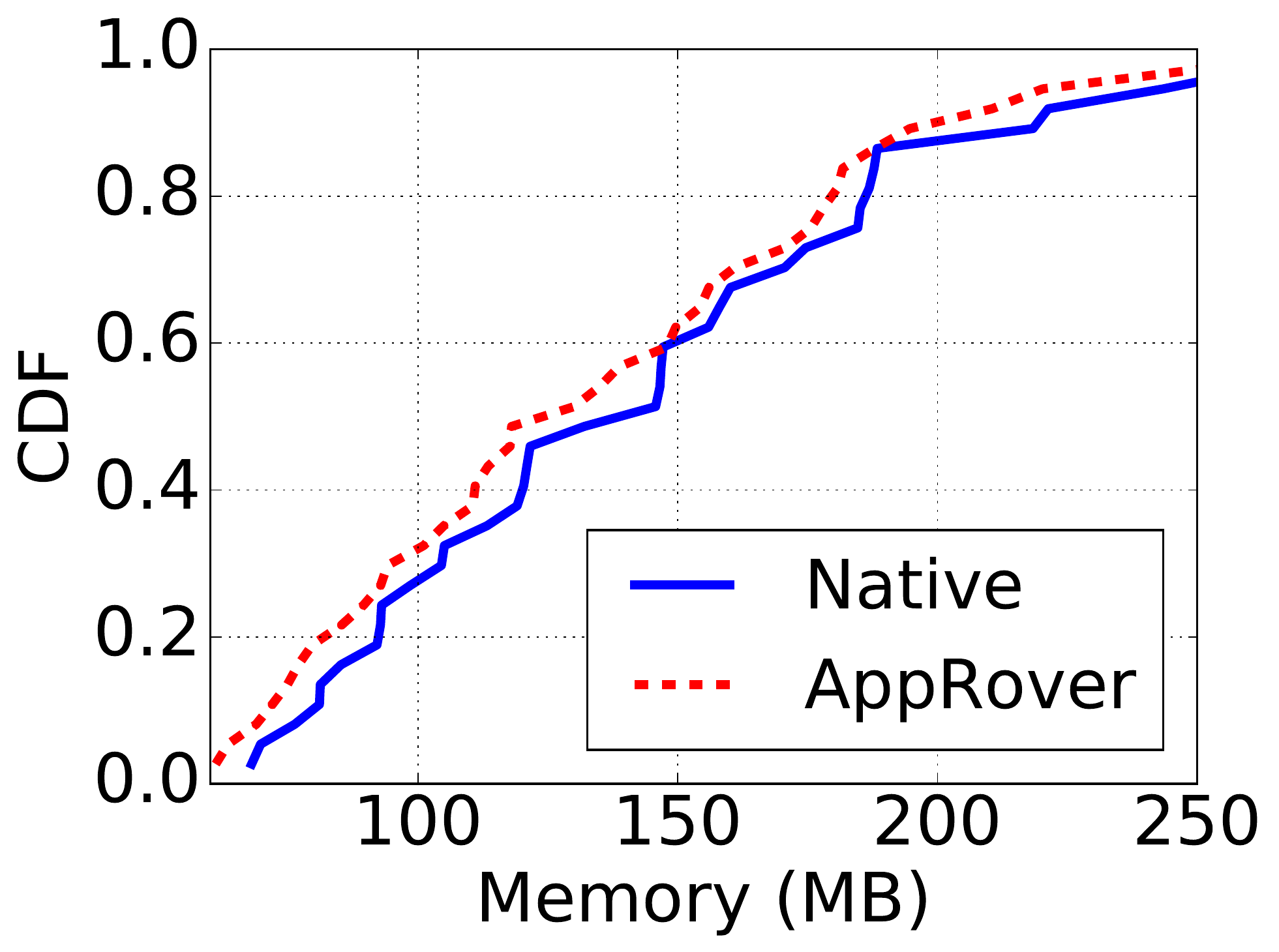}}
 \vspace{-1em}
  \caption{Runtime Performance}
  \vspace{-1em}
  \label{fig:evaluation}
\end{figure*}

Finally, we evaluate the runtime performance of the \nickName{} client using the 50 apps obtained in Section~\ref{subsec:robustness}. 
We use a Nexus 6 (3GB RAM, 32GB Rom, Quad-core 2.7 GHz) smartphone running Android 7.1.1 as our test device.

We measure the overhead when launching the base bundles with \nickName{} client compared to the overhead of launching the original app directly on the same platform. For each app, we choose the activities that are successor nodes of the home activity on the activity transition graph as the feature activities for testing, and open them with the \nickName{} client. For fair comparisons, we then launch the same activity in the original app directly. We compare the performance in terms of two metrics: activity load time~\cite{launch_time} and memory (RAM) usage. For a feature activity, we measure the performance on both cold start (the feature bundle is not installed on the device) and warm start (the feature bundle has been loaded previously). The cold start includes downloading the feature bundle, merging the resources, and loading the code. The warm start only needs to load the local code.  Figure~\ref{fig:evaluation} shows the results.

\noindent \textbf{Performance of launching the base bundle}. Figure~\ref{fig:time_on_launch} shows that launching the base bundle in the \nickName{} client performs even better than directly launching the original app. The base bundle contains only a part of code and resources extracted from the original app, and thus the launching process is more efficient. The median value for launching an app in the \nickName{} is about 654 ms, compared to 734 ms for the original app (about 10.9\% saving in the launching time). We also measure the memory usage as shown in Figure~\ref{fig:launch_memory}. We find that launching the base bundle requires less memory compared to launching the original app directly. In the median case, launching an app in \nickName{} client can save 6.5\% memory usage. It is worth to note that, the benefit from the reduced code base is somehow compromised by the overhead introduced by the virtualization layer of the \nickName{} client.
The reason is that our current implementation depends on \textit{VritualApp}, which loads some unnecessary libraries and incurs extra overhead. However, such an overhead can be further mitigated by implementing our approach at the kernel space.

\noindent \textbf{Performance of launching feature bundles}. We also find that the code start process of a feature bundle takes a much longer time than the warm start process. The bottleneck lies in the merging process. The \nickName{} client needs to download and merge the resources from the feature bundle into the base bundle, which involves is a time-consuming recompression process~\cite{incrementalupdate}.  Figure~\ref{fig:time_on_merge_feature_bundle} shows that the median value for the time used by the \nickName client to merge the resources extracted from a feature bundle is 2404.9 ms. However, we can eliminate such a cost by prefetching the feature bundle in the background according to others' access frequencies~\cite{WWW2010Zhou}. In practice, end users do not have to wait for these time-consuming processes on warm start, because the later visit needs to only load the code on local device directly. Although the \nickName{} client introduces a virtualization layer that may incur extra overhead and latency, a decomposed app with a smaller size can be still launched faster. Figure~\ref{fig:feature_load_time_warm} shows that opening a feature activity in the \nickName client takes a comparable time for the warm start compared to opening the same activity in the original app directly. For some cases, the activity load time can be even reduced with \nickName{}. The median value of the time for opening a feature activity in the \nickName{} client is about 216ms, compared to 270ms in the Android OS (saving about \textbf{20\%} of theopening time).

Figure~\ref{fig:load_memory} shows the memory overhead when opening an activity in the decomposed app and the original app, respectively. We just measure the memory usage after the activity is launched, because the memory varies during the launching process. In the median case, launching an app in \nickName{} can save \textbf{10.7\%} of the used memory.

\section{Discussion}
\label{sec:discussion}
In this section, we discuss some issues that could affect the effectiveness of our approach.

\noindent \textbf{Missing code and resources}. Due to the limitation of static program  analysis, the decomposed bundles may miss some code or resources. Therefore, \nickName{} uses the record-and-replay technique to iteratively launch each bundle, performs actions in the corresponding activity to explore the  behaviors, and collects the information of missing code and resources that are later included in the bundle. However, it is difficult to exhaustively explore all the behaviors in an activity. As a result, there may still exist some missing code and resources when the unexplored behaviors are triggered. To address this issue, we plan to  develop a runtime complementary mechanism in the \nickName{} client to retrieve the missing code and resources.

\noindent \textbf{App-level virtualization}. Although \nickName{} indeed brings improvement compared to directly running an app, there are some more space of optimization. Our current implementation is based on the \textit{VirtualApp} container. VirtualApp works at the user space, and requires a lot of computation resources to virtualize various system services. As a result, the overhead of the VirtualApp may potentially compromise the performance improvement gained by \nickName{}. Theoretically, \nickName{} could be implemented at the kernel space of Android OS  to reduce the overhead. Indeed, modifying the  kernel may not be as deployable and practical as the current \nickName{} client implemented upon VirtualApp, unless the modification can be adopted and deployed into the Android OS.

\noindent \textbf{Consistency with app-version  update}. \nickName{} decomposes an app into multiple bundles, and may break the existing app-updating mechanism where users download and re-install a new APK file to update an app. To address this issue, \nickName{} can download the latest bundles that contain the changed code and resources when a new version of the app is released. Actually, such an approach could bring benefits to the existing app-updating mechanism by enabling to update apps incrementally at runtime.

\noindent \textbf{Security concerns}. \nickName{} loads feature bundles in the way of dynamic code loading based on the ~\textit{DexClassLoader} provided by the Android system. However Poeplau et al.~\cite{NDSS14Poeplau} identified severe vulnerabilities (e.g., remote code injection) related to incorrect usage of dynamic code loading. To address this issue, we can check the integrity of each bundle before it is dynamically loaded~\cite{ACSAC15Falsina}.

\noindent \textbf{Potential additional cost}. We have demonstrated that  \nickName{} can save both local storage and memory usage compared to directly launching the original app. Ideally, we can suppose that the energy drain can be also reduced. Indeed, other potential additional cost should also be discussed. Requesting the feature bundles requires the data transfer, and the cost can vary according to the network connections and bandwidth. Hence, we cannot simply guarantee that \nickName{} can always gain benefits. However, given that the feature bundles are usually non-frequently visited pages, \nickName{} still takes values in most cases.

\section{Related Work}\
\label{sec:related_work}

\textbf{Software bloat}. McGrenere~\cite{McGrenereCHI2000} conducted an user study to unveil the software bloat problem of complex PC software. On average, users only used 27\% of functions, and they differed in which functions were unused. A. Akiki et al.~\cite{Akiki13aEICS}\cite{Akiki13bEICS} provided tools to simply and customize user interfaces of complex enterprise softwares so that end users were aware of only those UI feature-set. Tencent officially provides Mini Programs that are essentially Web applications to provide native app-like experiences for low frequency interactions~\cite{miniprogram}. Google provides Instant App~\cite{instantapp} with extra development efforts to reconstruct existing Android apps, but users have to download full APKs if their need other features. We offer an approach to decompose existing Android apps automatically, and users can visit those temporarily removed activities on demand.

\textbf{Decompostion}. Gui et al.~\cite{ICSE15Gui} and Liu et al.~\cite{mobisys2015Liu} statically analyzed apps to find the usage of ad libraries and rewrote bytecode for privilege de-escalation or better user experience. Carzaniga et al.~\cite{Carzaniga14ICSE} and Jelschen et al.~\cite{Jelschen12CSMR} detected and removed enery-wasting code in Android apps with the knowledge of energy-inefficiency patterns. Rubinov et al.~\cite{ICSE2016Rubinov}, Zhang et al.~\cite{OOPSLA2012Zhang}\cite{FCS12Zhang}, Wang et al.~\cite{Splash12Wang} and Liu et al.~\cite{TOIT17Liu} focused on extracting and offloading code to remote servers or trusted environments. Huang et al.~\cite{huangASE2017} proposed a static technique to remove code elements that were relevant to user specified unwanted UI elements in Android apps. Our approach considers to remove the infrequently visited activities and provides a mechanism to visit these removed activities on demand.

\textbf{Virtualization}. Chen et al.~\cite{Chen15TC} and Sun et al.~\cite{Sun2013ICCT} enforced isolation of apps and data to mitigate security risks by lightweight Android virtualization based on container technology. LBE Tech~\cite{multidroid} developed an virtualization engine called MutliDroid that empowered a user to run apps in the virtual operating system the same way the app ran in the original Android operating environment. Ki et al.~\cite{Mobisys17Ki} proposed a system called Reptor that enabled developers to modify and instrument Android platform API call behavior with app-layer API virtualization. Our approach offers an app-level virtualization space to take over the management of apps' lifecycle on non-rooted devices.

\section{Conclusion}
\label{sec:conclusion}

In this paper, we have proposed a service-oriented approach called \nickName{} to decompose Android Apps with app-level virtualization. \nickName{} allows developers to decompose an app into multiple bundles, and visit activities on demand. \nickName{} does not require any platform-level changes, and can run decomposed apps without installation. Our evaluations show that \nickName{} can save the initial download size of an app, and achieve better runtime performance than the native Android system when running the same app due to the reduced app size. 

\bibliography{ref}

\end{document}